\documentclass[12pt,preprint]{aastex}

\begin{document}
\title{H-Band and Spectroscopic Properties of Abell 1644}
\author{Aaron W. Tustin, Margaret J. Geller, and Scott
J. Kenyon\altaffilmark{1}}
\affil{Harvard-Smithsonian Center for Astrophysics, 60 Garden Street,
Cambridge, MA 02138}
\altaffiltext{1}{Visiting Astronomer, Cerro Tololo Inter-American
Observatory.  CTIO is operated by AURA, Inc. under contract to the
National Science Foundation.}  
\email{atustin@cfa.harvard.edu, mgeller@cfa.harvard.edu,
skenyon@cfa.harvard.edu}
\and
\author{Antonaldo Diaferio} 
\affil{Universit\`{a} degli Studi di Torino, Dipartimento di Fisica
Generale Amedeo Avogadro, Torino, Italy}
\email{diaferio@ph.unito.it}

\begin{abstract}

We discuss {\it H}-band (1.65 $\mu$m) near-infrared photometry of the
central $9\:h^{-2}$ Mpc$^2$ of Abell 1644 to a limiting $M_H\sim
M^*_H+3$ (throughout this paper $H_0=100\:h$ km s$^{-1}$ Mpc$^{-1}$).
There are 861 galaxies in the photometric survey region.  We also
measured radial velocities of 155 galaxies; 141 of these are cluster
members within $2.44\:h^{-1}$ Mpc of the cluster center.  The
completeness limit of the spectroscopic survey is $H\sim13$ ($M_H\sim
M^*_H$).  The cluster velocity dispersion of $\sigma\sim1000$ km
s$^{-1}$~remains constant out to the limiting radius.  We find no
evidence for substructure in the cluster.  The cluster mass within
$R=2.4\:h^{-1}$ Mpc is $7.6\pm1.3\times10^{14}\:h^{-1}\:M_\odot$.

We compute the cluster luminosity function; the Schechter parameters
$\alpha=-1.14\pm0.08$ and $M^*_H=-24.3\pm0.2$ (with $h=0.5$) agree
well with other {\it H}-band luminosity functions.  From the virial
theorem and the caustic method we compute one of the first
mass-to-light ratios at {\it H}; the result is
$M/L_H=82-127\:h\:M_\odot/L_\odot$ within $1.5\:h^{-1}$ Mpc.  This
ratio corresponds to $374-579\:h\:M_\odot/L_\odot$ at {\it R}.  The
agreement of our IR measurement with previous $M/L$ determinations
indicates that at low redshift dust and young stellar populations may
produce only negligible systematic errors in optical mass-to-light
ratios.

\end{abstract}


\section{INTRODUCTION}

Observations of galaxy clusters are a useful tool for constraining
models of structure formation and of determining cosmological
parameters.  Cluster mass-to-light ratios, traditionally measured at
optical wavelengths, are important in estimating $\Omega_0$
(c.f. Bahcall 1977; Bahcall, Lubin, \& Dorman 1995; Carlberg et
al. 1996; and references therein).  The luminosity functions (LFs) of
clusters can also be compared to field LFs to investigate the effects
of galaxy evolution in the cluster environment.

Infrared observations of clusters have several advantages over optical
measurements.  IR emission is less affected than visible light by
absorption and scattering due to dust, which may be interstellar or
intracluster \citep{zwi62,boy88} in origin.  Galactic infrared
emission is also dominated by the older stellar population and is
relatively insensitive to short-lived blue stars and changes in the
star formation rate \citep{bru93}.  Infrared radiation should
therefore be a good tracer of the total stellar mass of the cluster.

Until now, the small fields of view of infrared CCD arrays have
impeded deep large-area near-infrared photometry of clusters.  For
example, \citet{and00} measure the {\it H}-band LF of the Coma cluster
to a faint limiting magnitude ($M^*_H+6$), but in a projected area of
2\% of the size of our survey.  The 2MASS project surveys the
entire sky but is almost 2 mag shallower than our data at {\it H}.
Cluster surveys at {\it K} are slightly more common
\citep{dep99,tre98,bar98}, but most suffer from similar limitations.

The {\it H}-band data we discuss here, covering a projected area of
$9\:h^{-2}$ Mpc$^2$ and reaching $M_H\sim M^*_H+3$, make Abell 1644
(A1644) one of the best-studied clusters at near-infrared wavelengths.
We constrain the IR LF and compare its shape and parameters to those
for other cluster and field surveys.  The addition of extensive
spectroscopy enables us to comment on several kinematic issues.  We
examine the properties of emission and absorption galaxies, test for
substructure, and compute the infrared mass-to-light ratio of the
cluster.  This measurement, $M/L_H\sim82-127\:h\:M_\odot/L_\odot$,
is consistent with observed optical values and is one of the only
existing cluster $M/L$ determinations in the infrared.

In Section~\ref{observations} we describe our photometric and
spectroscopic observations.  Section~\ref{kinematics} contains an
analysis of the kinematic properties of the cluster.  In
Section~\ref{photometry} we compute the LF and mass-to-light ratio,
and we summarize in Section~\ref{discussion}.

\section{OBSERVATIONS} \label{observations}

Abell 1644 is a nearby rich cluster with a cD galaxy ($\alpha=12^{\rm
h}57^{\rm m}11^{\rm s}.6$,
$\delta=-17^{\circ}24^{\prime}34^{\prime\prime}.4$ J2000, $cz =
14,233\pm33$ km s$^{-1}$) at its center.  Our {\it H}-band photometric
observations cover a $3\times3\:h^{-2}$ Mpc$^2$ region surrounding the
cluster core to limiting $H\sim16$.  We also measured redshifts for
155 galaxies within 2.44 $h^{-1}$ Mpc of the cluster center.  The
combined sample of 123 cluster members with both spectral and
photometric data is complete to $H\sim13$.

\subsection{Photometry}

We obtained {\it H}-band photometry covering a
$36^{\prime}\times36^{\prime}$ region centered on $\alpha=12^{\rm
h}57^{\rm m}35^{\rm s}$ $\delta=-17^{\circ}27^{\prime}21^{\prime
\prime}$ on 1997 April 16-19.  We used the CIRIM infrared camera on
the 1.5 m telescope at Cerro Tololo Inter-American Observatory (CTIO).
The scale for the $256\times256$ array is $1^{\prime \prime}.137$
pixel$^{-1}$, and the duration of each exposure was $6\times10^{\rm
s}$.  We operated the instrument in a mapping mode which acquired
pairs of dithered images offset by $10^{\prime\prime}$.  We then
covered the survey region with a tiling of dithered pairs which
overlap by $40^{\prime\prime}$.  In addition, we surveyed two
off-cluster fields of sizes 0.22 deg$^2$ and 0.04 deg$^2$ to ascertain
background statistics.

On the nights of 1999 April 25-29 and 1999 May 26 we used the
Stelircam IR camera on the 1.2 m telescope at Fred L. Whipple
Observatory (FLWO) on Mt. Hopkins, Arizona to expand our infrared
survey.  We operated the $256\times256$ instrument (1$^{\prime
\prime}$.2 pixel$^{-1}$) in a mapping mode similar to that used at
CTIO.  Of the six nights at FLWO, two were not photometric.  On the
remaining nights we gathered data over a 0.99 deg$^2$ region bordering
the CTIO observations (bringing our total coverage in the cluster to
1.35 deg$^2$), and we increased the sizes of our off-fields to 0.41
deg$^2$ and 0.44 deg$^2$.  We also imaged a small area in the center
of the cluster, overlapping the CTIO data, to assess systematic
differences between the two surveys.

\begin{figure}
\plotone{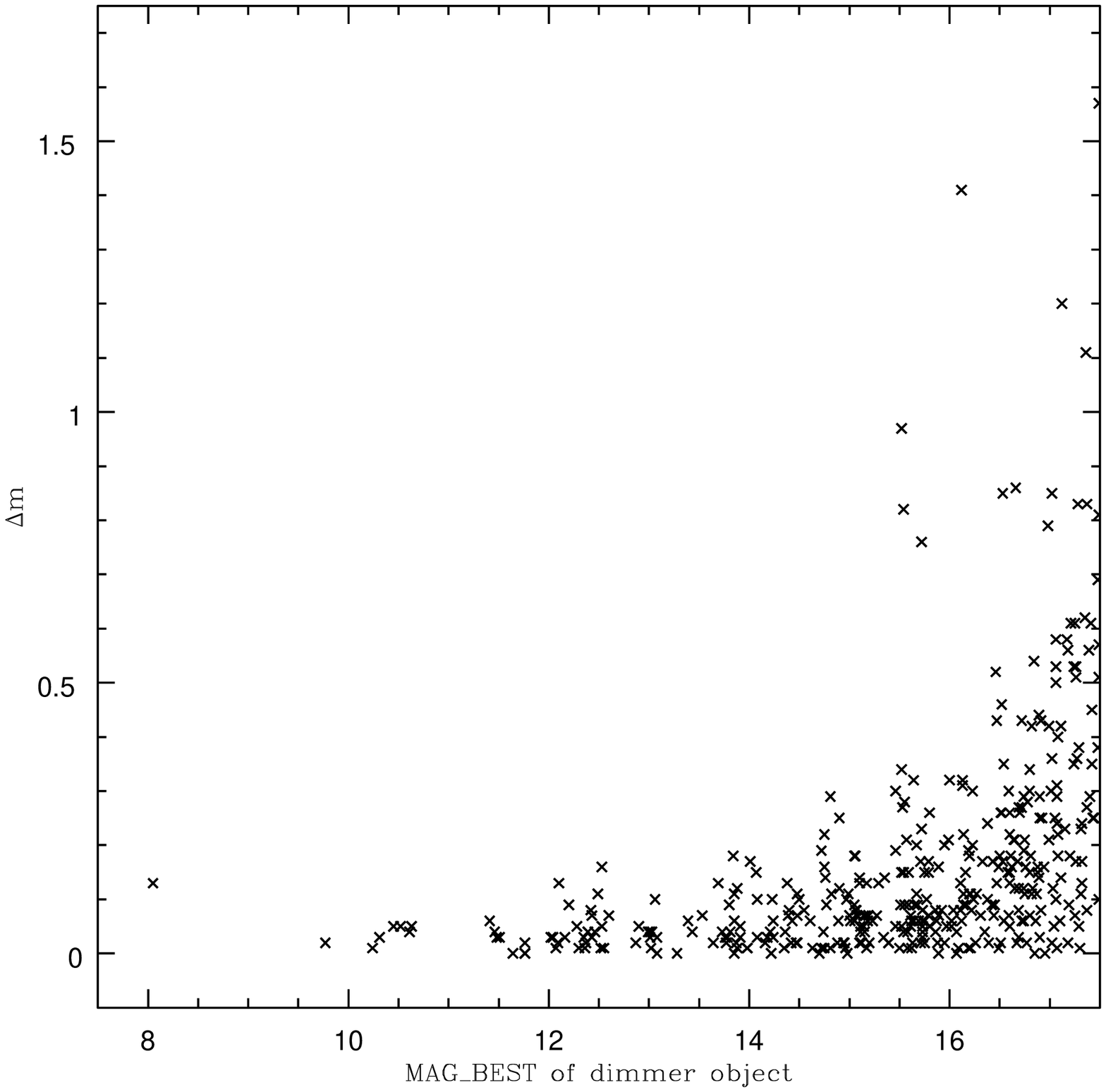}
\caption{Magnitude difference versus dimmer magnitude for objects
observed in two successive dithered frames on 1997 April 17.  At
MAG\_BEST $\sim16$ the rms magnitude difference is 0.12 mag.}
\label{fig-mdiff}
\end{figure}

For each photometric night, we first used NOAO IRAF\footnote{IRAF is
distributed by the National Optical Astronomy Observatory, which is
operated by the Association of Universities for Research in Astronomy,
Inc. under contract to the National Science Foundation.} commands to
flat-field and sky-subtract each frame.  We calculated the zero point
of each night's magnitude scale by fitting a linear solution to
several standard star images and correcting each object frame for
airmass.  We then used the SExtractor software package for source
detection and photometry.  In addition to the usual aperture and
isophotal magnitudes, SExtractor calculates a composite MAG\_BEST.
MAG\_BEST is useful in crowded fields and is equal to the Kron
magnitude \citep{kro80} if no neighbor can bias the Kron aperture
magnitude by more than 10\%.  Otherwise it is the corrected isophotal
magnitude \citep{ber96}.  We adopt MAG\_BEST for our subsequent data
analysis.  To demonstrate the consistency of the photometry, we plot
the magnitude differences as a function of MAG\_BEST for objects
observed in successive dithered frames on 1997 April 17
(Figure~\ref{fig-mdiff}).  The rms error in the photometry is $\Delta
m_{rms}\sim0.12$ mag at $H=16$.

The SExtractor software also includes an algorithm to separate stars
and galaxies based on morphology.  Each object identified has
a value CLASS\_STAR which ranges from zero for galaxies to one
for stars.  Ambiguous objects lie somewhere in between.  We examine
the images manually and find that CLASS\_STAR accurately separates
stars and galaxies brighter than $H\sim14.5$; fainter objects are
indistinguishable by eye in these images.  

An examination of the CTIO and FLWO data in the overlap region reveals
that the seeing is better in the CTIO images.  However, there is
little systematic difference in the photometry.  The average
magnitude difference between objects observed at FLWO and CTIO is
$\overline{\Delta m}=0.04$ mag, and the rms scatter is $\Delta
m_{rms}=0.15$ mag.  The SExtractor star/galaxy
separation gives consistent results for objects observed at both
sites.  In the overlap region, 93\% of the objects with $H<14.5$
identified as galaxies in the FLWO data are also classified as
galaxies in the CTIO data.

Table~\ref{table1} shows photometric data for the 861 galaxies in our
infrared survey.  For $H<13$ the galaxy spectroscopic survey is
complete; for $13<H<16$ we list all objects with ${\rm
CLASS\_STAR}<0.5$.  Columns (1) and (2) give the right ascension and
declination of each galaxy; column (3), the apparent $H$ magnitude
(MAG\_BEST); and column (4), the redshift if measured.

\subsection{Spectroscopy}

We acquired optical spectra for 155 galaxies in the region of Abell
1644 during 2000 January-June with FAST, a high throughput slit
spectrometer mounted on the 1.5 m telescope at FLWO.  The FAST setup
includes a 300 groove mm$^{-1}$ grating and 3$^{\prime \prime}$ slit,
and yields spectra which cover 3600-7600\AA~at a resolution of
$\sim6$\AA.  Although \citet{dre8a} measured 103 redshifts in the
region of Abell 1644, we re-acquired spectra of 91 of their objects to
assure uniformity throughout our data set and to assess the spectral
type of the galaxies.  The remaining twelve DS galaxies are either too
faint for FAST redshift measurements or are outside our survey region.
We selected 64 additional targets on the basis of the DS catalog,
digitized sky survey images, and our own infrared survey.

We obtained a reduced spectrum for each object using IRAF.  We made
bias, flat-field, and illumination corrections to each frame and then
used the IRAF XCSAO task to obtain redshifts via cross-correlation
with standard emission and absorption templates.  The resulting
spectroscopic sample is complete to $H\sim13$ in the region where we
have {\it H}-band photometry.  We tested the differences between the
FAST and DS observations for the 91 common galaxies.  The FAST
velocities are systematically lower by $\sim39$ km s$^{-1}$, and the
rms velocity difference is $\Delta v_{rms}=80$ km s$^{-1}$.  These
offsets are consistent with the estimated velocity errors of roughly
30 km s$^{-1}$ (FAST) and 45 km s$^{-1}$ (DS).

Table~\ref{table2} lists properties of the 155 galaxies with measured
redshifts.  Columns (1) and (2) give the right ascension and
declination; column (3), the radial velocity; column (4), the error on
the radial velocity; column (5), the spectral type (either emission or
absorption as discussed in Section~\ref{spectral}); column (6),
MAG\_BEST (if available); and column (7), the reference number from the 
DS catalog for galaxies in the DS list.

\subsection{Cluster Sample}

\begin{figure}
\plotone{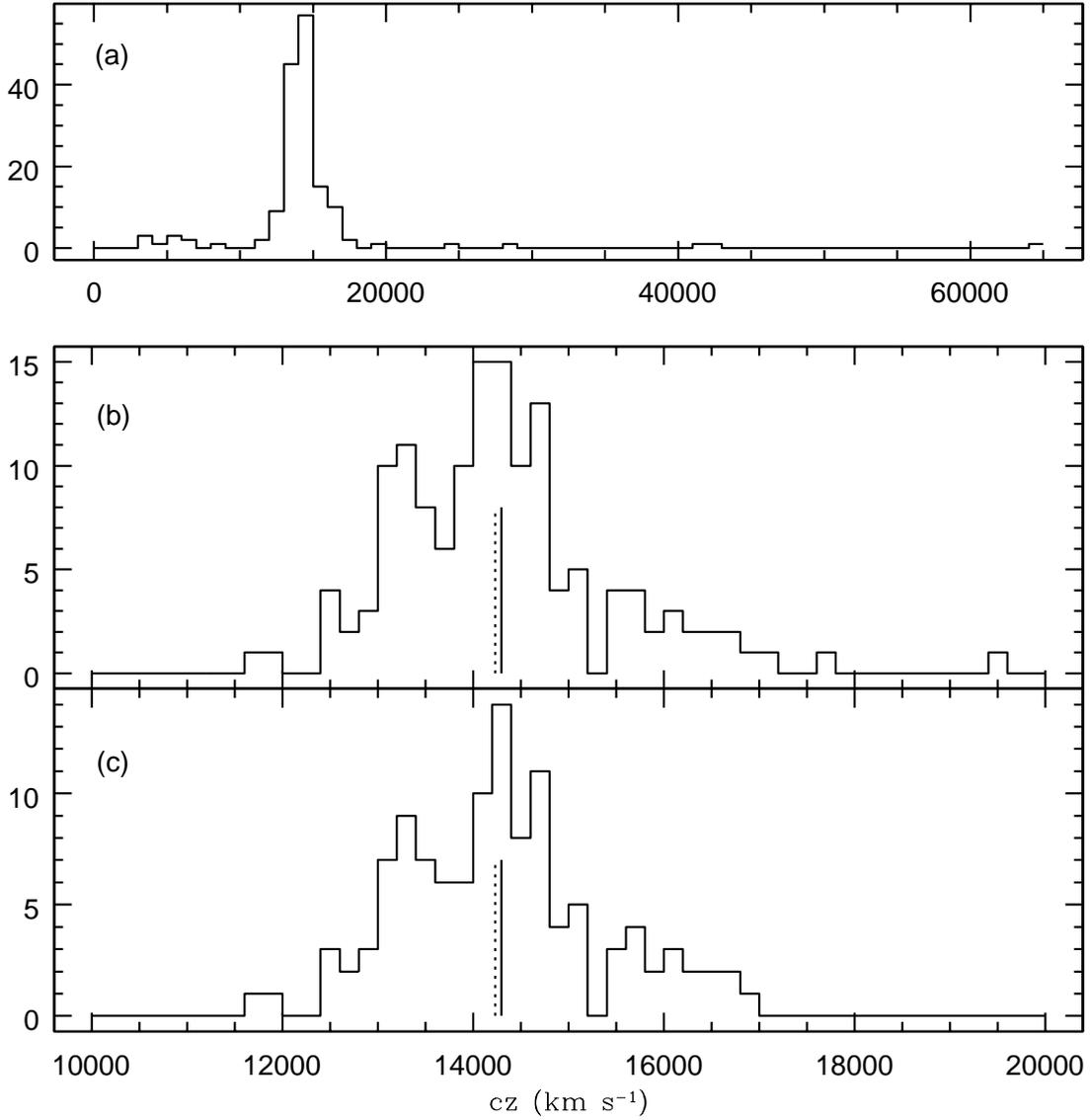}
\caption{a) Velocity distribution of all 155 galaxies observed with
FAST.  b) Cluster sample of 141 galaxies in the range
$10,000<cz<20,000$ km s$^{-1}$.  c) 123 cluster members for which {\it
H}-band photometry is available.  In the bottom two plots, the dotted
vertical line is the velocity of the cD while the solid line is the
mean radial velocity of the cluster.}
\label{fig-vhist}
\end{figure}

Figure~\ref{fig-vhist}a shows the measured velocity distribution.
A1644 is remarkably isolated in redshift space.  A few background
galaxies are visible at redshifts up to $cz\sim65,000$ km s$^{-1}$.
There is also evidence of foreground structure in the range 2500 $< cz
<$ 10,000 km s$^{-1}$.  Figure~\ref{fig-vhist}b shows the 141 cluster
members with redshifts between 10,000 and 20,000 km s$^{-1}$.

The mean radial velocity of the cluster is $c\overline{z}=14,295\pm93$
km s$^{-1}$~with a dispersion of $\sigma=1108^{+73}_{-61}$ km
s$^{-1}$, computed at the 68\% confidence limit as in \citet{dan80}.
The radial velocity of the cD ($14,233\pm33$ km s$^{-1}$) is
coincident with the overall cluster mean.  The cluster sample is
isolated from foreground and background galaxies by velocity gaps of
3244 km s$^{-1}$~on the low end and 4589 km s$^{-1}$~on the high end.
The sizes of these gaps are $2.9\sigma$ and $4.1\sigma$, respectively.
There are 123 cluster members within the region of infrared
photometry; Figure~\ref{fig-vhist}c shows the velocity distribution of
these galaxies.

\section{KINEMATICS} \label{kinematics}

\subsection{Emission and Absorption Galaxies} \label{spectral}

\begin{figure}
\plotone{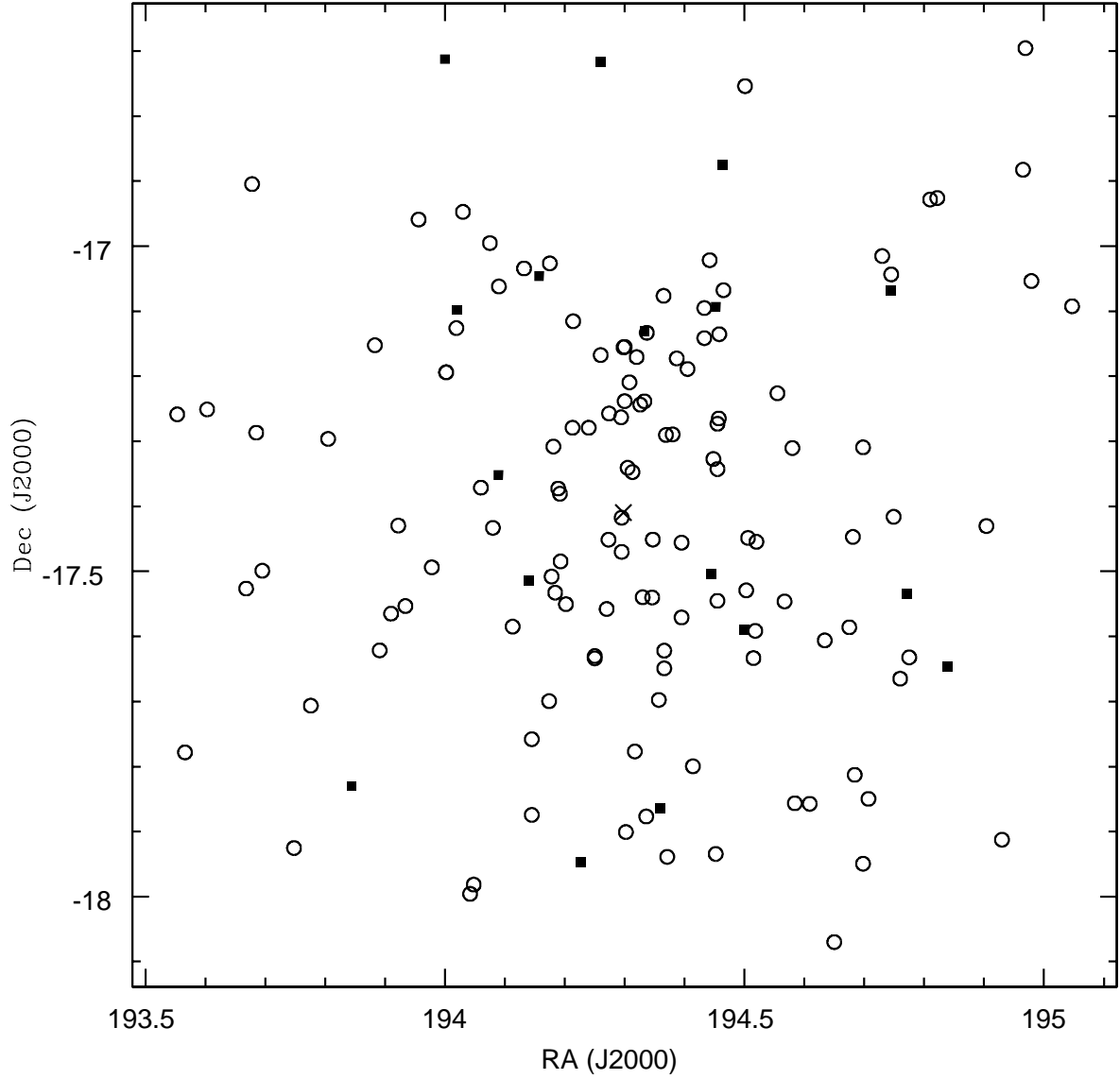}
\caption{Sky positions of the 141 cluster members.  Right ascension
and declination are in decimal degrees.  Open circles represent Abs
galaxies and filled squares show Em systems.  The cD, an absorption
galaxy, is indicated by a cross.}
\label{fig-pos}
\end{figure}

We separate the galaxies into emission (Em) and absorption (Abs)
systems.  Emission galaxies are those objects with a larger R value
when cross-correlated with the emission template; all other galaxies
are classified as absorption.  Figure~\ref{fig-pos} shows the positions
of the 141 cluster members according to spectral type.  There are 18
emission and 123 absorption systems.  The absorption galaxies are more
centrally condensed; the median projected distances of the Abs and Em
samples from the cD galaxy are $20^{\prime}.1$ and $27^{\prime}.5$,
respectively.  This difference is only marginally statistically
significant; a rank-sum test applied to the two samples gives a 97\%
confidence that the Em and Abs systems do not come from the same
parent spatial distribution.

\begin{figure}
\plotone{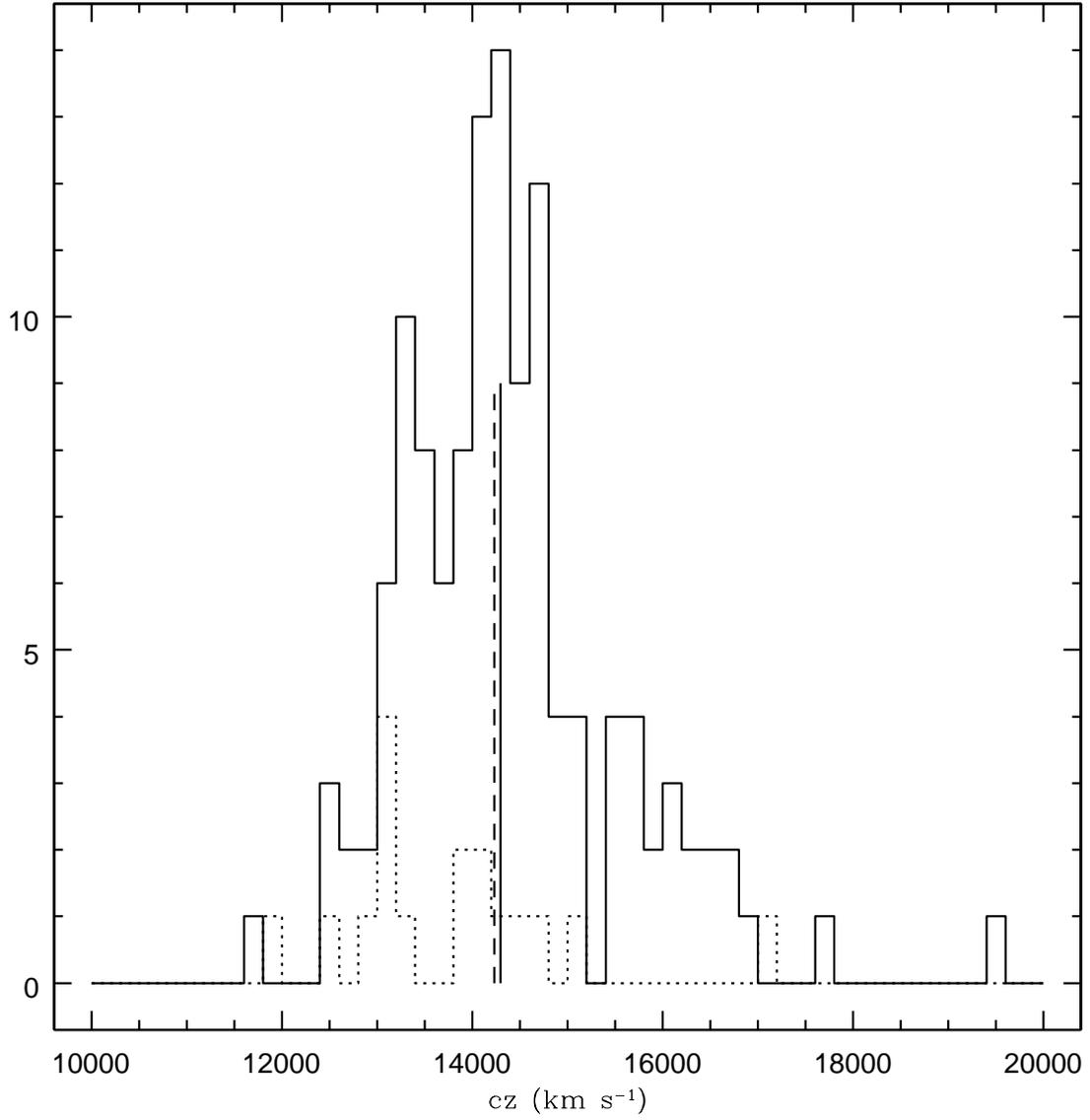}
\caption{Velocity distribution of the cluster sample separated by
spectral type.  Solid histogram: 123 Abs galaxies.  Dotted histogram:
18 Em galaxies.  Solid and dashed vertical lines indicate the mean
velocity of the cluster and the velocity of the cD, respectively.}
\label{fig-veldist}
\end{figure}

Figure~\ref{fig-veldist} shows the radial velocity distributions of the Abs
and Em samples.  The absorption galaxies have
$c\overline{z}=14,363\pm98$ km s$^{-1}$~and $\sigma=1095^{+77}_{-64}$
km s$^{-1}$.  The dispersion of the Em sample
($\sigma=1123^{+268}_{-156}$ km s$^{-1}$) is comparable, but the mean
velocity of $c\overline{z}=13,800\pm273$ km s$^{-1}$~is lower.
However, a rank-sum test gives only a 96\% confidence that the Em and
Abs galaxies follow different velocity distributions.

\subsection{Substructure}

\citet{dre8b} define a test parameter ${\delta}$ which measures the
deviation of a subgroup's local mean velocity and dispersion,
$\overline{v}_{local}$ and ${\sigma}_{local}$, from that of the
cluster's overall values ($\overline{v}$ and $\sigma$):
\begin{equation}
\delta^{2} = \frac{n}{\sigma^{2}}[(\overline{v}_{local} -
\overline{v})^{2} + (\sigma_{local} - \sigma)^{2}],
\end{equation}
where $n$ is the number of galaxies in the subgroup.  The cumulative
deviation is $\Delta=\sum^N_{i=1}\delta_i$, where $N$ is the number of
cluster members and for each galaxy we calculate $\delta_i$ using the
galaxy and its $n-1$ nearest neighbors.  For a Gaussian velocity
distribution with only random fluctuations, $\Delta\sim N$; the
presence of substructure can cause $\Delta$ to be significantly higher
than $N$.
     
\citet{dre8b} analyze 92 members of A1644 with $n=11$ and find
evidence for substructure in the southeastern quadrant of the cluster.
To test the significance of this structure, Dressler and Shectman
randomly shuffle the velocities of the galaxies to create roughly 1000
model clusters.  They calculate the value of $\Delta$ for each model
and find that only 2.7\% of the Monte Carlo clusters have
$\Delta\ge\Delta_{\rm observed}$.

\begin{figure}
\plotone{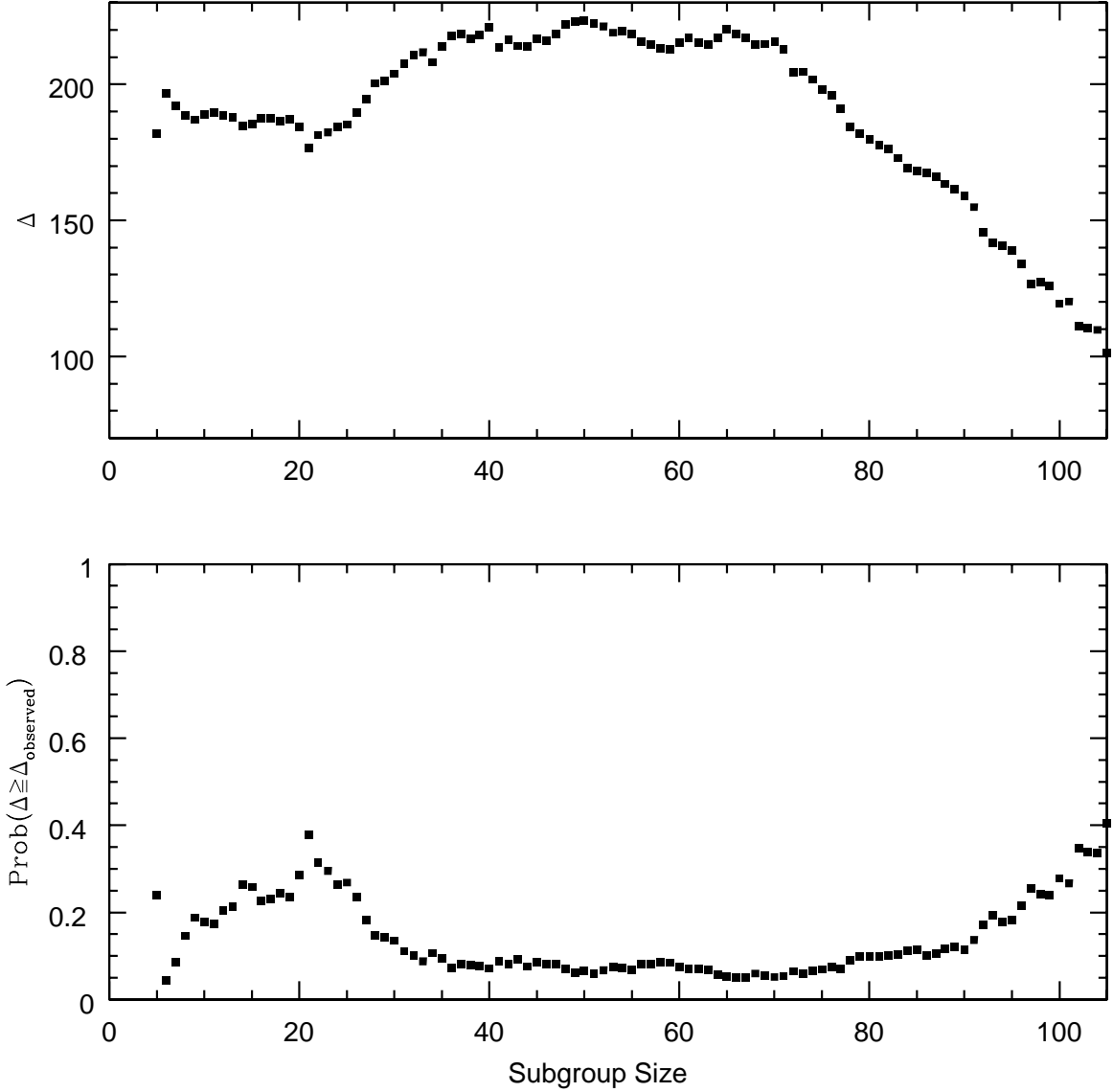}
\caption{Top: Value of the Dressler-Shectman statistic $\Delta$ for
varying subgroup size $n$.  Bottom: Probability that
$\Delta\ge\Delta_{\rm observed}$, based on 5000 Monte Carlo models for
each value of $n$.  With $n=11$ our analysis does not confirm the
detection of substructure by \citet{dre8b}.}
\label{fig-ds}
\end{figure}

We duplicate the Dressler and Shectman analysis for our larger sample
of 141 cluster members.  Allowing $n$ to range from five to 105, we
calculate the $\Delta$ statistic for each subgroup size.  We also
create 5000 Monte Carlo models for each value of $n$ and compute the
probability that $\Delta\ge\Delta_{\rm observed}$.  Figure~\ref{fig-ds}
shows the results.

With $n=11$, we find that $\Delta$ is greater than or equal to
$\Delta_{\rm observed}$ in 17.5\% of the simulated clusters.  Thus our
larger sample, when compared to the Dressler and Shectman catalog, no
longer provides evidence for substructure on this scale.  The best
evidence for substructure occurs with $n=6$ and $n=67$.  However,
these detections are not statistically significant, with probabilities
of $\Delta\ge\Delta_{\rm observed}$ of 4.3\% and 5.0\%, respectively.

\subsection{Velocity Dispersion Profile}

\begin{figure}
\plotone{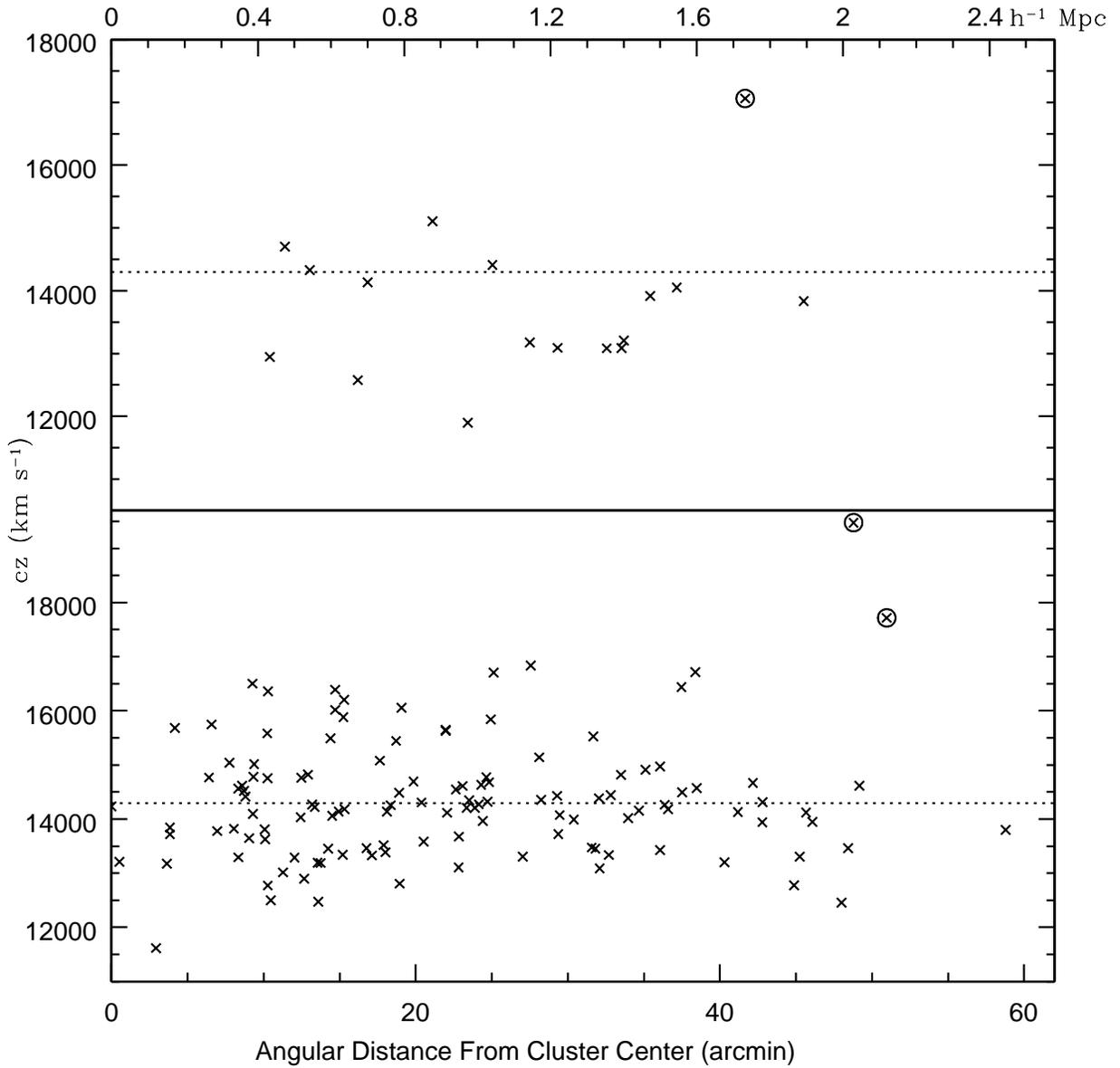}
\caption{Radial velocity profile of Em (top) and Abs (bottom)
galaxies.  The horizontal dotted line shows the mean velocity of the
cluster.  There are three circled galaxies which may be outliers.}
\label{fig-vrad}
\end{figure}

\begin{figure}
\plotone{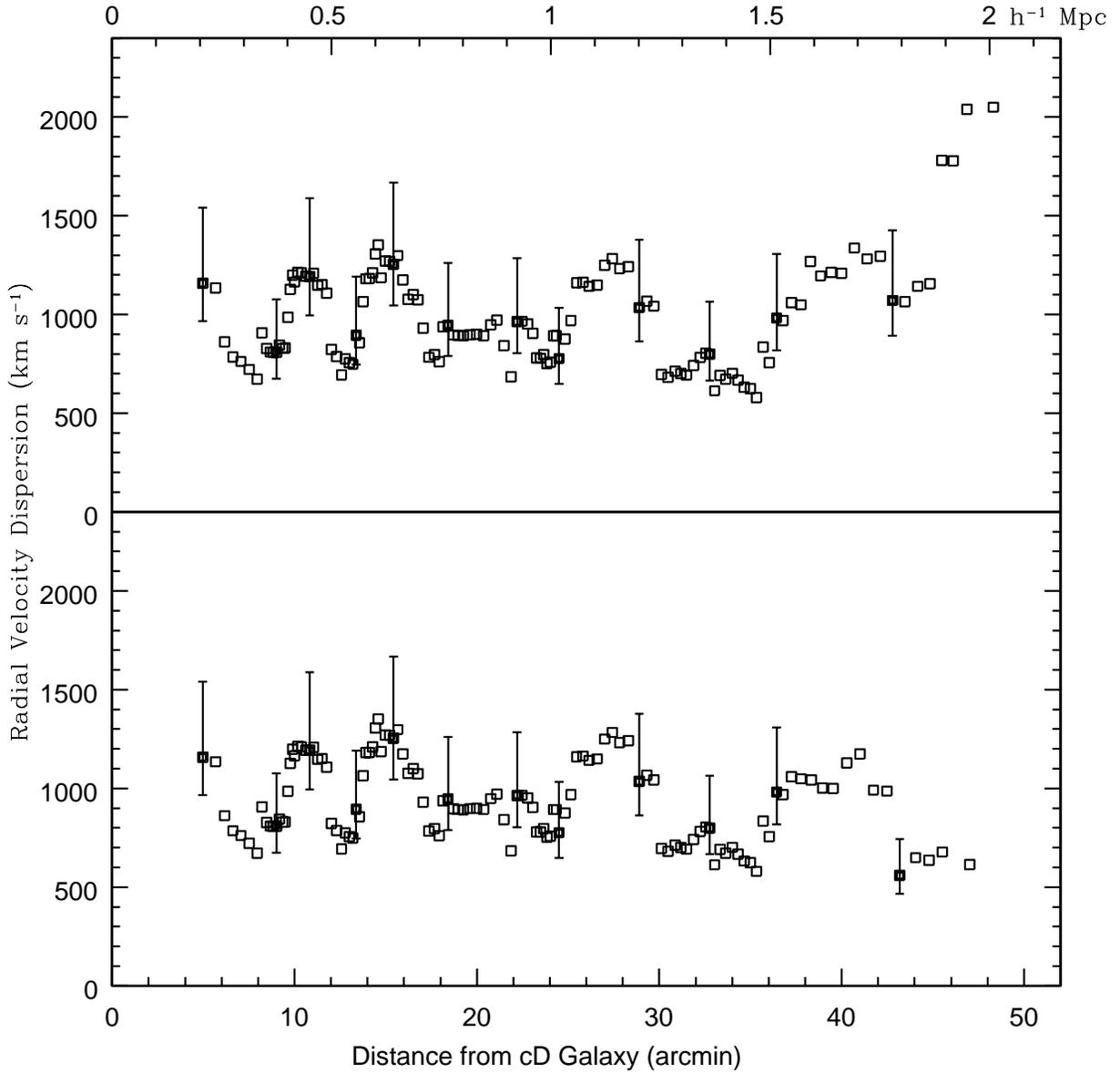}
\caption{Velocity dispersion for groups of 11 galaxies moving outward
from the cD.  Error bars show 68\% confidence levels for uncorrelated
samples.  Top: Includes all 141 galaxies with 10,000 km
s$^{-1}<cz<20,000$ km s$^{-1}$.  Bottom: Excludes the three
high-velocity outliers circled in Figure~\ref{fig-vrad}.}
\label{fig-disp}
\end{figure}

Figure~\ref{fig-vrad} shows an azimuthally averaged radial profile
of the cluster velocity distribution for both Abs and Em galaxies.  We
compute the velocity dispersion of the cluster as a function of
distance from the cD galaxy and plot the results in
Figure~\ref{fig-disp}.  We choose the cD as the geometric center of
the cluster because of the lack of evidence for substructure.  Moving
outward from the cD, we calculate the velocity dispersion of
successive groups of 11 galaxies arranged in order of increasing
radius.  The error bars in the plot represent the 68\% confidence level 
for uncorrelated points.

The velocity dispersion of the cluster remains roughly constant for
$R\lesssim1.9h^{-1}$ Mpc; at this radius it nearly doubles.  The
increase at large radius is extremely sensitive to the presence of
three high-velocity galaxies circled in Figure~\ref{fig-vrad}.  When we
exclude these galaxies (Figure~\ref{fig-disp}, bottom plot) the
velocity dispersion at $R\sim1.9\:h^{-1}$ Mpc decreases by a factor of
three.  We suggest that the three high-velocity galaxies may be outliers
and that $\sigma$ is roughly constant as a function of radius.
Treating these objects as background is consistent with the results of
our caustic analysis (Section~\ref{mass}), and also with
\citet{dre8b}, who select $cz\lesssim18,000$ km s$^{-1}$~as a
criterion for membership in A1644.

\subsection{Cluster Mass} \label{mass}
     
We calculate the mass of A1644 in two ways.  Our first mass estimate
depends on the ``caustic'' technique of \citet{dia97}.  A hierarchical
clustering model predicts the existence of two caustic curves with
amplitude $A(R)$ approximately equal to the escape velocity from the
cluster at radius $R$.  \citet{dia99} shows that $A(R)$ is related to
the mass of the cluster interior to $R$:
\begin{equation}
\label{eqn-caustic}
GM_{est}(<R)=\frac{1}{2}\int^R_0A^2(x)dx.
\end{equation}

\begin{figure}
\plotone{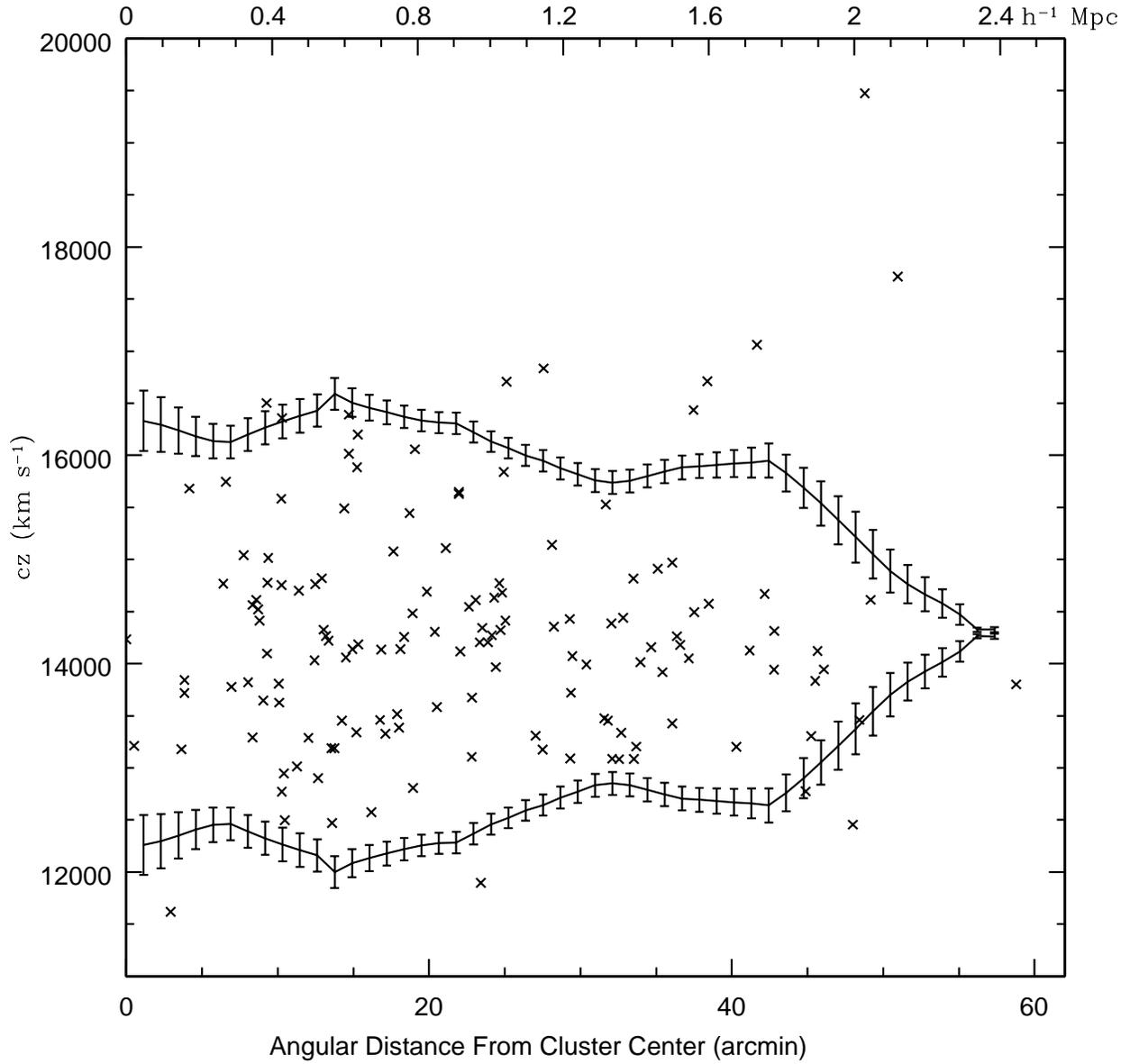}
\caption{Caustics and cluster members in position-redshift space.  The
caustic curves $A(R)$ are shown as solid lines with $1\sigma$ error
bars.}
\label{fig-caust}
\end{figure}

We use the techniques of \citet{dia99} with smoothing parameter $q=25$
to calculate the caustics of A1644 (Figure~\ref{fig-caust}).  Of the 141
presumed cluster members, 127 lie within the caustics; the fourteen
remaining galaxies may in fact be outliers. Equation~\ref{eqn-caustic}
automatically excludes these outliers from the caustic mass
determination.

Our second mass estimate uses a virial estimator \citep{bin87}:
\begin{equation}
\label{virial-eqn}
M_{est} = \frac{3 \pi N}{2 G}\frac{\sum^{N}_{i=1}
v^{2}_{p,i}}{\sum^{N}_{i=1} \sum_{j<i} |{\bf {\it R_{i}}} - {\bf {\it
R_{j}}}|^{-1}},
\end{equation}
where $v_{p,i}$ is the radial velocity of each galaxy with respect to
the cluster mean and {\bf $R_i$} is the galaxy's position relative to
the cD.  This estimator assumes that the galaxies are embedded in a
diffuse dark matter distribution, and that the spatial arrangement of
galaxies traces the dark matter.

\begin{figure}
\plotone{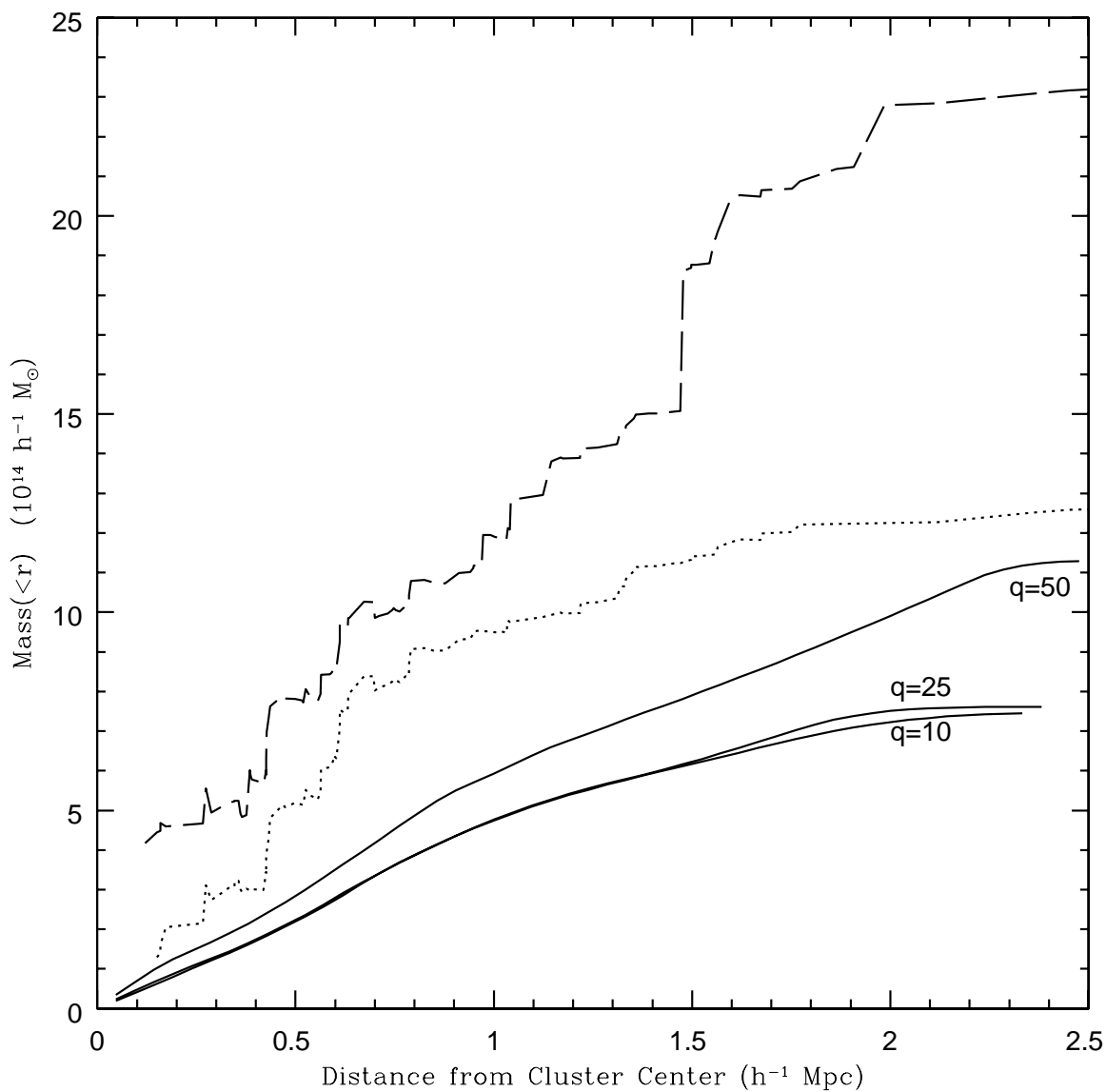}
\caption{Virial and caustic masses illustrating the inherent
uncertainties in each technique.  Solid lines: caustic mass calculated
with smoothing parameter {\it q}=10, 25, and 50.  Dashed line: virial
mass including 141 galaxies with 10,000 km s$^{-1}<cz<20,000$ km
s$^{-1}$.  Dotted line: virial mass using only galaxies which lie
within the {\it q}=25 caustics.}
\label{fig-massdiff}
\end{figure}

The virial mass is very sensitive to outliers.  In
Figure~\ref{fig-massdiff} we first plot $M_{virial}$ using the 141
galaxies with 10,000 km s$^{-1}<cz<20,000$ km s$^{-1}$.  The virial
mass decreases by a factor of two or more, even at large radius, when
we include only those galaxies which lie inside the caustics.  For
comparison, we plot $M_{caustic}$ for three different values of the
smoothing parameter; the caustic mass is much more robust.

Equation~\ref{virial-eqn} overestimates the true value of $M$ because
we have not subtracted the virial surface term \citep{the86}.  When
the surface term $C(r)$ is included, the corrected virial mass is
\begin{equation}
\label{correctedvir-eqn}
M_{cv}(<b)=M_{est}\{1-C(b)\}=M_{est}\biggl\{1-4\pi
b^3\frac{\rho(b)}{\int^b_0 4\pi
r^2\rho(r)dr}\biggl[\frac{\sigma_r(b)}{\sigma(<b)}\biggl]^2\biggl\},
\end{equation}
where $\rho(r)$ is the radial density distribution, $\sigma_r$ is the
radial component of the velocity dispersion, and $\sigma(<b)$ is the
integrated velocity dispersion within the limiting radius $b$
\citep{gir98}.

Calculation of the surface term requires knowledge of both the
velocity dispersion profile and the density distribution.  While
$\sigma(r)$ can be constrained from the data, the density profile is
unknown and is, in fact, the very quantity we are attempting to
correct via equation~\ref{correctedvir-eqn}.  To approximate $\rho(r)$
we fit a Navarro, Frenk, \& White (1995, 1996, 1997, hereafter NFW)
mass profile to $M_{caustic}$.  The NFW functional forms of the mass
and density profiles are
\begin{equation}
\label{nfw_mass}
M(<r)=4\pi \delta_c \rho_c
r_c^3\biggl[\log(1+r/r_c)-\frac{r/r_c}{1+r/r_c}\biggl]
\end{equation}
and
\begin{equation}
\label{nfw_rho}
\rho(r)=\frac{\delta_c \rho_c}{(r/r_c)(1+r/r_c)^2},
\end{equation}
where $\delta_c$ and $r_c$ are model parameters and $\rho_c$ is the
critical density of the universe.  We fit equation~\ref{nfw_mass} to
the caustic mass profile and substitute the best-fit parameters into
equation~\ref{nfw_rho} to recover the density distribution.  We then
use the fitted $\rho(r)$ to estimate the virial surface term via
equation~\ref{correctedvir-eqn}.

To estimate $C(r)$ one must also assume a value of the anisotropy
parameter $\beta(r)=1-\sigma_\theta^2/\sigma_r^2$.  \citet{gir98} show
that clusters with a flat velocity dispersion profile are best
described by models with $\beta(r)=0$.  We adopt this value of
$\beta(r)$ and compute the virial surface term.  The overall trend of
the surface term is decreasing with radius.  However, $C(r)$
fluctuates on small scales because of its dependence on $\sigma(r)$.
At $r=1.5\:h^{-1}$ Mpc, $C(r)\sim0.2\pm0.1$.

We use a statistical ``jackknife'' procedure \citep{dia83} to
calculate errors in the virial mass, $\Delta M/M\sim2-4\%$ for
$R=0.5-1.5\:h^{-1}$ Mpc.  Systematic sources of error in the virial
mass dwarf this formal jackknife estimate; the surface term correction
is highly uncertain due to the assumptions regarding $\rho(r)$ and
$\beta(r)$.  In addition, the virial estimate depends strongly on the
particular galaxy sample.

\begin{figure}
\plotone{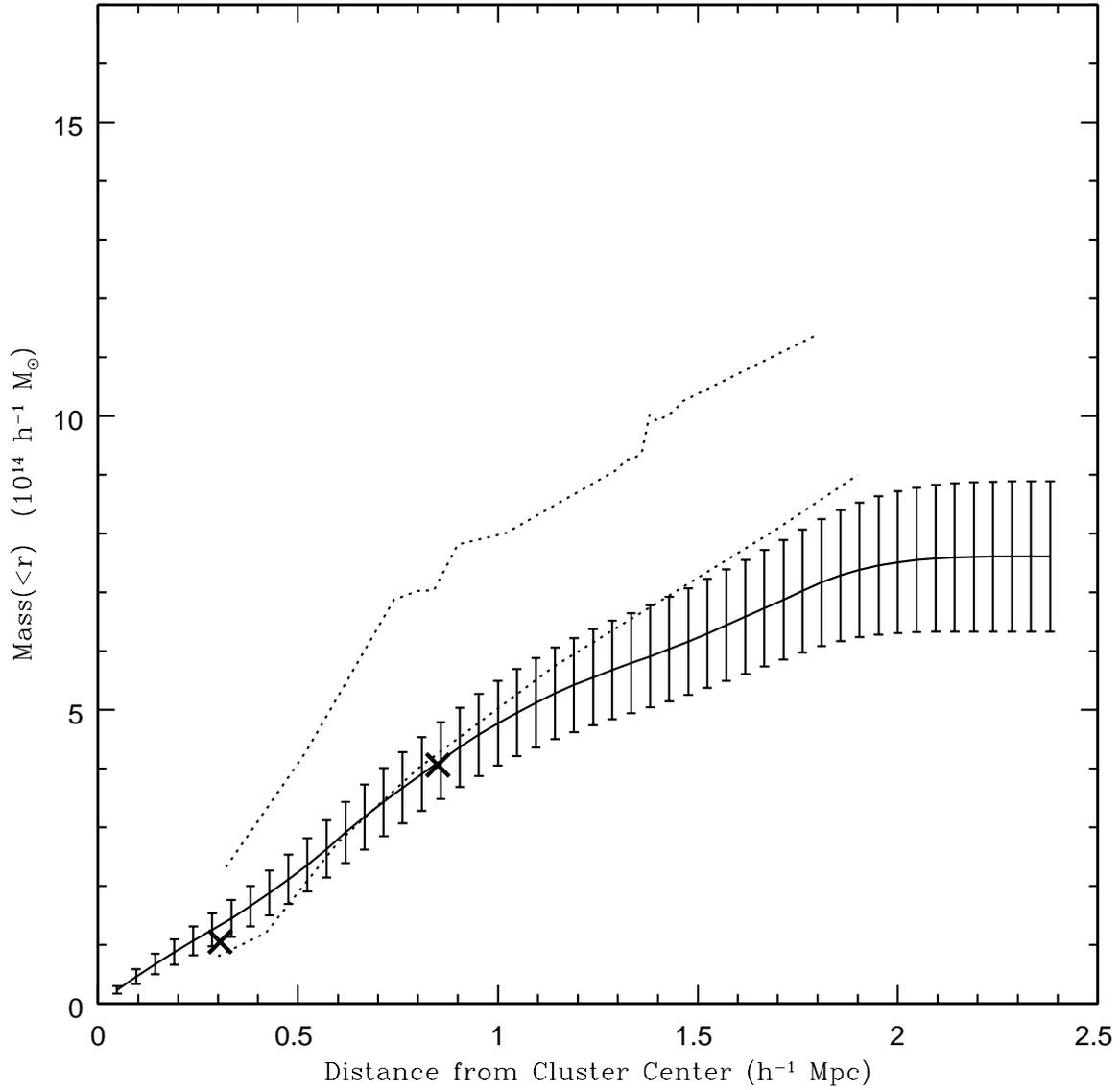}
\caption{Radial mass profile of A1644.  Solid line represents the
caustic estimate; dotted lines show upper and lower envelopes of the
virial estimate with surface term correction.  Crosses show the X-ray
determinations from \citet{ett97}.}
\label{fig-massprof}
\end{figure}

Figure~\ref{fig-massprof} shows our final estimates of the caustic and
virial mass profiles.  For consistency, we base both estimates on the
same set of 127 galaxies enclosed by the caustics.  Errors in
$M_{caustic}$ represent uncertainty in the location of the caustics.
The virial mass incorporates the surface term correction.  We show
$M_{virial}$ as an envelope whose maximum and minimum values signify
the range of masses consistent with the jackknife errors and the
uncertainty in $C(r)$.  With our adopted errors, the caustic
and virial mass estimators are consistent at the $2\sigma$ level at
all radii.

Figure~\ref{fig-massprof} also includes two X-ray mass determinations
from \citet{ett97}.  They arrive at mass estimates for A1644 based on
{\it Einstein Observatory} IPC X-ray data.  Their deprojection
analysis yields $M_{dpr}=1.05\times10^{14}\:h^{-1}\:M_\odot$ within
$0.305\:h^{-1}$ Mpc.  They then extrapolate the deprojection mass to
$R_{500}$, the radius where $\rho/\rho_{crit}=500$; the extrapolated
mass is $M_{500}=4.06\times10^{14}\:h^{-1}\:M_\odot$ within
$0.85\:h^{-1}$ Mpc.  Our caustic mass estimate is consistent with the
X-ray data at both radii.

\section{PHOTOMETRIC PROPERTIES} \label{photometry}

\subsection{Luminosity Function}

\begin{figure}
\plotone{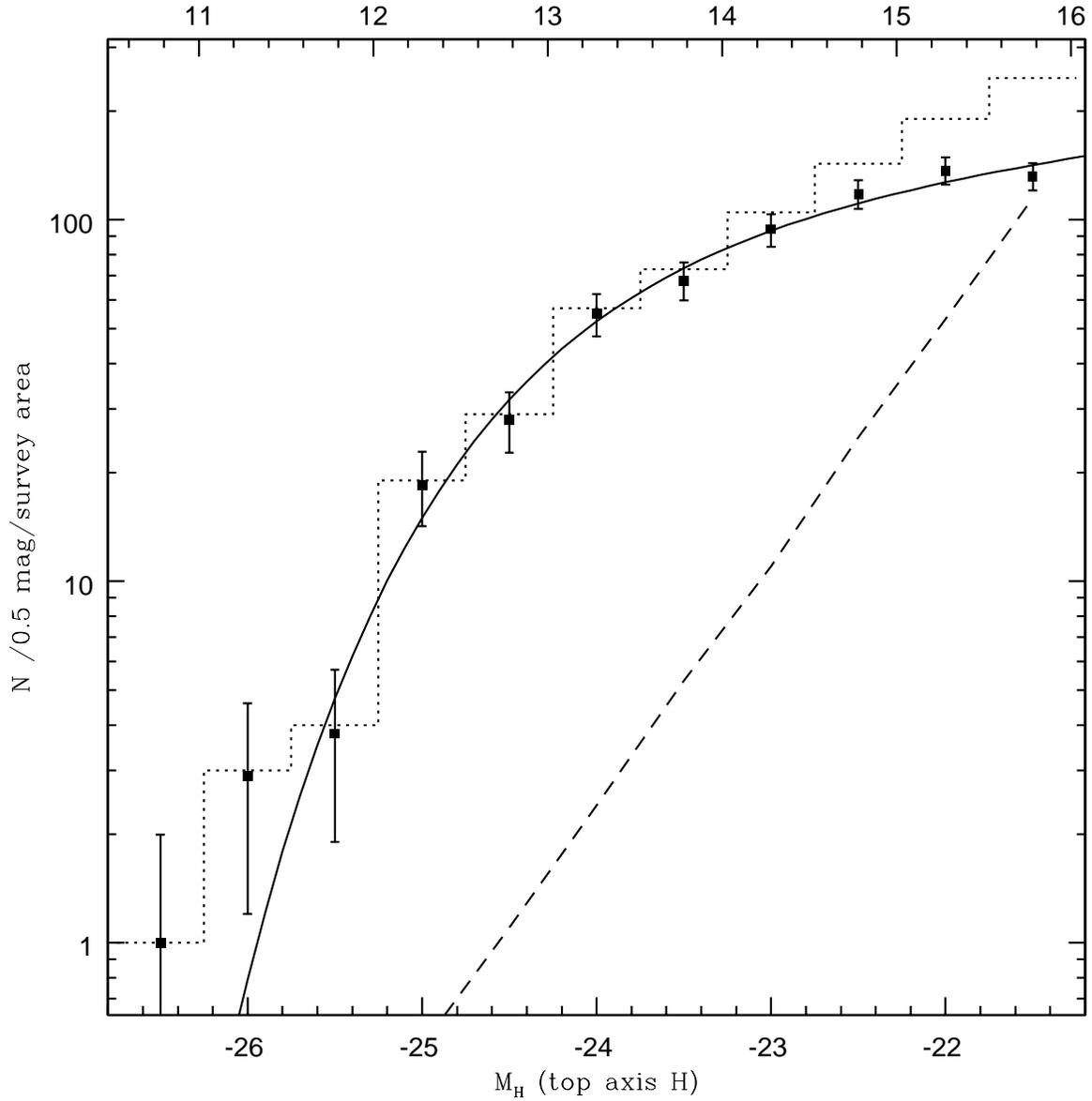}
\caption{Differential {\it H}-band luminosity function.  Dotted
histogram: all objects classified as galaxies by the SExtractor
CLASS\_STAR parameter.  Dashed line: background counts modeled as a
10$^{0.67m}$ power law.  Filled squares with error bars: number of
cluster members in each bin after background subtraction.  Solid
curve: best-fit Schechter function.}
\label{fig-lf}
\end{figure}

Figure~\ref{fig-lf} shows the differential {\it H}-band luminosity
function of A1644.  In the bins with complete spectroscopy ($H<13$)
redshifts determine cluster membership.  At the faint end we estimate
the number of field galaxies because we do not have spectra.  We model
the number of field galaxies in each magnitude bin by a power law of
the form
\begin{equation}
\label{power-law}
N (m_1<m<m_2) = C\int^{m_2}_{m_1}10^{0.67m} dm,
\end{equation}
where $C$ is a normalization constant.  We assume the background slope
of 0.67 determined by \citet{gar93} from a compilation of {\it K}-band
field surveys.  As our best estimate of the background, we normalize
the power law by integrating equation~\ref{power-law} from $H=-\infty$
to $H=14$ and equating the left hand side with the number of field
galaxies in the direction of the cluster; there are $5\pm2$ field
galaxies deg$^{-2}$ with $H<14$ in our photometric region.  Our
off-cluster survey contains $39\pm6$ bright galaxies per square degree,
and Gardner et al. (1993) and \citet{szo98} observe roughly $13\pm3$
and $23\pm6$ galaxies deg$^{-2}$ in this magnitude range.  Although
these field counts are statistically consistent with our estimate, the
large variations in the normalization can produce a rising, flat, or
declining faint end of the A1644 cluster LF.

After subtracting the background fit normalized by our redshift survey
from the total number of faint galaxies identified by the SExtractor
CLASS\_STAR parameter, we characterize the data with a Schechter
luminosity function.  We fit to the \citet{sch76} form of the
luminosity function
\begin{equation}
\label{lf-eqn}
n_e(L)dL=n^*(L/L^*)^\alpha\exp(-L/L^*)d(L/L^*),
\end{equation}
where $\alpha$ is the faint-end slope, $L^*$ is a characteristic
luminosity, and $n^*$ is a normalization constant.  In terms of
absolute magnitude $M$ the luminosity function is
\begin{equation}
N_e(M)dM=k\:n^*\exp\{[-k(\alpha+1)(M-M^*)]-\exp[-k(M-M^*)]\}dM,
\end{equation}
where $k\equiv(\ln10/2.5)$ and $M^*$ is the absolute magnitude
corresponding to $L^*$ \citep{kas95}.  At the mean redshift of our
cluster $M_H=m_H-35.78+5\log h$.

We fit a Schechter function to our data by minimizing the quantity
\begin{equation}
\chi^2\equiv\sum\frac{[N(M_i)-N_e(M_i)]^2}{\sigma_i^2},
\end{equation}
where $N(M_i)$ and $N_e(M_i)$ are the observed and fitted numbers of
galaxies in the $i$th magnitude bin and $\sigma_i$ is the variance in
the $i$th measurement.  With respect to the maximum likelihood method
\citep{san79}, also commonly used to fit luminosity functions, a
$\chi^2$ fit has the disadvantage that the data must be binned, but it
is advantageous because it provides a measure of the goodness of the
fit.  We assume a Poissonian variance in the total number of galaxies
and in the number of background galaxies in each bin.

We determine the best-fit Schechter function (solid curve in
Figure~\ref{fig-lf}); the best-fit parameters are
$\alpha=-1.14\pm0.08$, $M^*=-24.3\pm0.2$, and $n^*=114\pm22$ (here and
in subsequent discussions of absolute magnitudes we take $h=0.5$).
The Schechter function is a good fit to the faint end of the LF and
slightly underpredicts the number of bright galaxies.  The minimum
value of $\chi^2$ is 6.3 with 8 degrees of freedom.

Our best-fit values of $\alpha$ and $M^*_H$ are in good agreement with
a survey of A1644 in the visible \citep{kas95}.  They fit a Schechter
LF (with $\alpha$ fixed at -1.25) to the {\it R}-band luminosity
distribution and find $M^*_R=-21.50$.  With an expected elliptical
color of $(R-H)\sim2.5$, the characteristic magnitude at {\it H} is
$M^*_H\sim-24.0$, which agrees very well with our measurement.

There are several {\it K}-band field luminosity functions.  Parameters
derived from these surveys span roughly $-1.0<\alpha<-1.3$ and
$-25.1<M^*_K<-23.6$ \citep{mob93,szo98,lov00}.  Our values for A1644
are consistent with this range, assuming $H-K=0.25$ as above.
Parameters derived from {\it K}-band studies of clusters are similar.
\citet{dep99} find $M^*_K\sim -24.9\pm0.5$ with $\alpha$ fixed at -0.9
for clusters in their $z=0.15$ bin.  A separate survey of five X-ray
luminous clusters finds a composite $\alpha=-1.38\pm0.24$
\citep{tre98}.  Our results are consistent with these low $z$ counts
for typical early-type galaxy color, but not with \citet{bar98}, who
find an average $M^*_K=-25.4\pm0.1$ for 10 clusters with
$0.31<z<0.56$.  The discrepancy could be due to luminosity evolution,
although Barger et al. claim that their $M^*_K$ is not significantly
brighter than the $z\sim0$ values.

At {\it H} fewer data are available.  An {\it H}-band study of a 0.1
deg$^2$ region of the Coma cluster finds $\alpha=-1.3\pm0.2$ and
$M^*_H=-24.6\pm1.0$, with a dip in the LF at $M_H=-22.2$ ($H\sim13.5$)
\citep{and00}.  We estimate the errors on their values of $\alpha$ and
$M^*_H$ by examining the 68\% confidence contours in their Figure 3.
An imaging study of A548 over a much larger area (0.9 deg$^2$) reveals
a similar dip at $H\sim13-13.5$ (S. J. Kenyon et al., in preparation).
The number of galaxies in our survey is too small to evaluate the
presence of a dip at $H\sim13.5$.  The counts in the $13<H<13.5$ and
$13.5<H<14$ bins ($55\pm7$ and $68\pm8$, respectively) are very
similar; the LF in this region is thus consistent with either a flat,
rising, or declining slope.

\subsection{Mass-to-Light Ratio} \label{ml}

We determine the total {\it H}-band luminosity of A1644 by
extrapolating the luminosity function.  For a magnitude-limited survey
described by a Schechter function the observed fraction of the total
light is given by $\Gamma(\alpha+2,L_{min}/L^*)/\Gamma(\alpha+2)$,
where $\Gamma(x,y)$ is the incomplete Gamma function and $L_{min}$ and
$L^*$ are the luminosities corresponding to the completeness limit and
$M^*$.  With a completeness limit of $H=16$, the fitted Schechter
parameters imply that we observe 90.6\% of the total light.  Our
photometric survey covers a square field 9 $h^{-2}$ Mpc$^2$ in area.
The photometry is therefore complete within $0.5(9\:h^{-2}$
Mpc$^2)^{1/2}=1.5\:h^{-1}$ Mpc of the center of the field.  Inside
this radius the total luminosity is
$L_H=7.6\pm0.4\times10^{12}\:h^{-2}\:L_{\odot}$; the quoted error in
$L_H$ reflects the uncertainty in the LF parameters.  

For comparison, we integrate the visible LF in a smaller 0.82 deg$^2$
region of A1644 \citep{kas95}.  Using their values of $\alpha$,
$M^*_R$, and $M^{lim}_R$, the total {\it R}-band light within 1.1
$h^{-1}$ Mpc is $L_R\sim1.1\times10^{12}\:h^{-2}\:L_\odot$.  Within
the same radius we compute the {\it H}-band luminosity and use typical
galactic colors and $(R-H)_\odot=0.85$ to convert {\it H} photometry
to {\it R}.  The result is
$L_H=5.1\pm0.3\times10^{12}\:h^{-2}\:L_\odot$ which implies
$L_R\sim1.12\pm0.06\times10^{12}\:h^{-2}\:L_\odot$, in excellent
agreement with Kashikawa et al.  The agreement implies that the optical
luminosity of A1644 is only slightly affected by factors such as dust
extinction and recent star formation activity.

\begin{figure}
\plotone{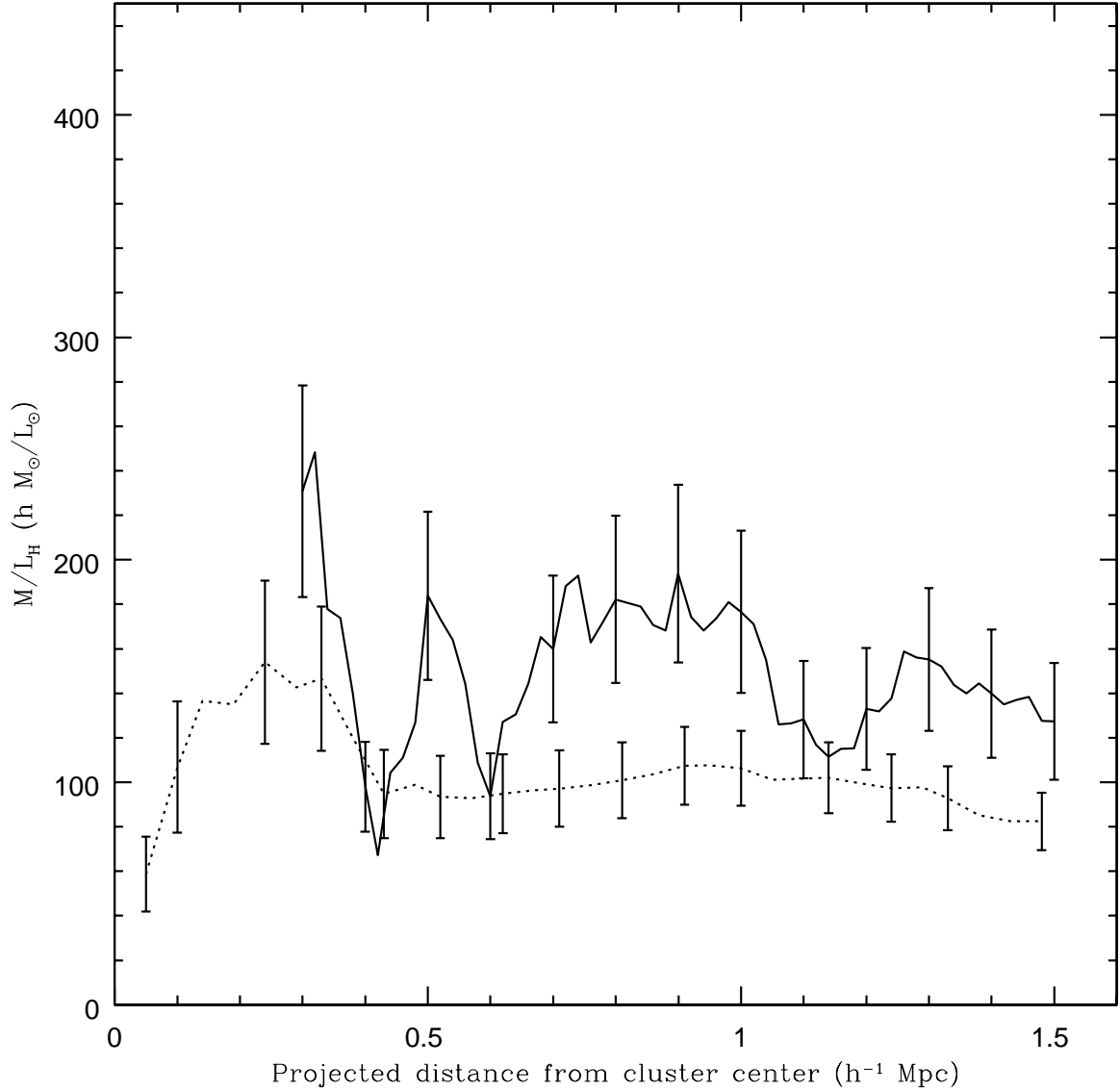}
\caption{Mass-to-light ratio as a function of projected radius.  Solid
and dotted lines represent $M/L_H$ computed with virial and caustic
mass estimates, respectively.  We show characteristic error bars at
several locations along each curve.}
\label{fig-ml}
\end{figure}

Figure~\ref{fig-ml} shows the radial $M/L_H$ profile.  The
mass-to-light ratio is flat for $R\gtrsim0.4\:h^{-1}$ Mpc, suggesting
that the dark matter fraction is constant outside this radius.  The
increased $M/L$ at $R\lesssim0.35\:h^{-1}$ Mpc is a result of a
depressed luminosity near the cluster core.  At the limiting
photometric radius of 1.5 $h^{-1}$ Mpc the caustic and virial
mass-to-light ratios are $M/L_H=82\pm13\:h\:M_\odot/L_\odot$ and
$M/L_H=127\pm26\:h\:M_\odot/L_\odot$.

For A1644, \citet{gir00} obtain $M/L_{B_j}\sim250\:h\:M_\odot/L_\odot$
within 1.5 $h^{-1}$ Mpc.  Transforming from $B_j$ to {\it H} gives
$M/L_H\sim40\:h\:M_\odot/L_\odot$ for their data.  This value is
roughly half of our caustic estimate.  Because our measured masses are
almost exactly the same, the difference in $M/L$ is due to their
larger luminosity.  Our luminosity determination is based on CCD
photometry and agrees well with the optical CCD photometry of
\citet{kas95}; Girardi et al.'s analysis is based on digitized and
calibrated survey plates from the COSMOS/UKST Southern Sky Object
Catalog and is discrepant with Kashikawa et al.

Cluster mass-to-light determinations at {\it H} are scarce.  We thus
transform to {\it R} and {\it V} using the $R-H$ colors above along with
$V-H\sim3$ for early-type galaxies and $(V-H)_{\odot}=1.37$.  The
mass-to-light ratios within 1.5 $h^{-1}$ Mpc are then equivalent to:
\begin{displaymath}
M_{caustic}/L_R=374\pm59\:h\:M_\odot/L_\odot
\end{displaymath}
\begin{displaymath}
M_{caustic}/L_V=369\pm59\:h\:M_\odot/L_\odot.
\end{displaymath}
\begin{displaymath}
M_{virial}/L_R=579\pm119\:h\:M_\odot/L_\odot
\end{displaymath}  
\begin{displaymath}
M_{virial}/L_V=572\pm117\:h\:M_\odot/L_\odot
\end{displaymath}

The caustic mass-to-light ratio is consistent with most previous
measurements of cluster $M/L$ ratios; typical values using virial
masses are $M/L_R\sim300\:h\:M_\odot/L_\odot$ \citep{car96,gir00},
with a range of $200-600\:h\:M_\odot/L_{\odot,R}$ \citep{dre78}.  Our
virial $M/L$ determination is consistent with the high end of this
range.  Estimates of $M/L$ using X-ray masses tend to be somewhat
lower. \citet{hra00} find $M/L_V=154-468\:h$ solar units for seven
nearby Abell clusters and one group, and {\it ROSAT} observations of
eleven groups and clusters yield $M/L_V=200-300\:h\:M_\odot/L_\odot$
\citep{dav95}.  

\section{DISCUSSION AND SUMMARY} \label{discussion}

We discuss a large, deep imaging survey of a galaxy
cluster in the near-infrared, covering a 1.35 deg$^2$ (9 $h^{-2}$
Mpc$^2$) region of A1644.  These data allow us to determine the IR LF
at {\it H} to roughly $M^*_H+3$, deeper than most previous
infrared surveys of comparable size.  We also acquired spectra for 155
galaxies (including $\sim141$ cluster members) and use these to compute
$M/L_H$ for the cluster.  The results are:

\begin{enumerate}
\item A grouping of cluster members by spectral type reveals evidence
at the 96-97\% confidence level for the segregation of emission and
absorption galaxies by both velocity and core radius.  Although only
marginally statistically significant, these detections of segregation
with respect to spectral type are in the same direction as the
morphological segregation observed in other clusters
\citep{ada98,mel77,biv92}.  In contrast with an earlier analysis by
\citet{dre8b}, there is no significant evidence for substructure on
any scale.  Our 50\% larger sample of cluster members is responsible
for the difference.

\item The velocity dispersion of the cluster varies little with radius
for $R\leq1.5\:h^{-1}$ Mpc.  Near the center, $\sigma$ is slightly
larger than at the periphery, provided that we exclude three
high-velocity galaxies at large clustercentric distance which may be
background.

\item The luminosity distribution is well-fit by a Schechter function
with $\alpha=-1.14\pm0.08$ and $M^*_H=-24.3\pm0.2$.  These parameters
agree with previous infrared LF measurements of clusters and field
galaxies.  In particular, an {\it H}-band determination of the Coma LF
\citep{and00} yields an identical slope and characteristic magnitude
to within the errors.  The shape of the A1644 LF is consistent with
several infrared cluster LFs \citep{tre98,bar98,dep99}.  However, due
to small number statistics and the relatively flat slope of the LF at
$H\sim13-13.5$ we are unable to rule out the presence of a dip similar
to the ones observed in optical and infrared cluster LFs
\citep{kor98,and00}.

\item We compute dynamical masses of A1644 based on the virial theorem
and on the use of caustics.  The caustic technique yields masses which
have smaller errors, are more robust, and agree better with X-ray
data.  These masses, combined with the {\it H}-band photometry, yield
an $M/L_H\sim82-127\:h\:M_\odot/L_\odot$ within $R=1.5\:h^{-1}$ Mpc.
This mass-to-light ratio is one of the first determinations in the
infrared, and corresponds to $\sim374-579\:h\:M_\odot/L_\odot$ at
visible wavelengths.  That the infrared $M/L$ agrees with previous
optical determinations suggests that in some clusters biases by
factors such as dust and short-lived star formation events are a small
effect compared with the errors in the determination of mass-to-light
ratios.
\end{enumerate}

\acknowledgments 

We thank the staff members of FLWO and CTIO for assistance with our
observations.  In particular, Mercedes G\'omez assisted with the
observations at CTIO, and the remote observers at Whipple Observatory
obtained all of the redshifts listed in Table~\ref{table2}.  We are
grateful to Susan Tokarz for reducing the FAST spectra.  We also thank
the referee for comments which increased the clarity of the paper.

\clearpage


\clearpage

\begin{deluxetable}{cccc}
\tabletypesize{\scriptsize} 
\tablecaption{Properties of Photometric
Sample\label{table1}} 
\tablewidth{0pt} 
\tablehead{ & & &
\colhead{$cz$} \\ \colhead{RA (J2000)} & \colhead{Dec. (J2000)} &
\colhead{$m_H$} & \colhead{km s$ ^{-1}$} \\ \colhead{(1)} &
\colhead{(2)} & \colhead{(3)} & \colhead{(4)} } 
\startdata 12 54 50.9 & -17 01 28.6 & 14.46 & \ldots \\ 12 54 51.4 &
-16 57 34.6 & 15.19 & \ldots \\ 12 54 57.9 & -17 00 59.0 & 15.15 &
\ldots \\ 12 55 06.2 & -17 42 23.0 & 13.97 & 14971 \\ 
\enddata
\tablecomments{The complete version of this table is in the electronic
edition of the Journal.  The printed edition contains only a sample.}
\end{deluxetable}

\clearpage

\begin{deluxetable}{ccccccc}
\tabletypesize{\scriptsize} 
\tablecaption{Properties of Spectroscopic Sample\label{table2}} 
\tablewidth{0pt} 
\tablehead{ & & \colhead{$cz$} & \colhead{$\sigma_{cz}$} & & & \colhead{DS} \\
\colhead{RA (J2000)} & \colhead{Dec. (J2000)} & \colhead{(km
s$^{-1}$)} & \colhead{(km s$^{-1}$)} & \colhead{Type} &
\colhead{$m_{H}$} & \colhead{Number}\\ \colhead{(1)} & \colhead{(2)} &
\colhead{(3)} & \colhead{(4)} & \colhead{(5)} & \colhead{(6)} &
\colhead{(7)} }
\startdata
12 58 35.9 & -18 04 11.8 & 12774 & 33 & Abs & 13.18 & 1\\
12 56 10.2 & -17 59 44.7 & 16712 & 36 & Abs & 14.20 & 2\\
12 56 54.5 & -17 56 51.2 & 13083 & 16 & Em & 12.16 & 6\\
12 52 20.9 & -17 39 16.0 & 13306 & 34 & Abs &\ldots & 9\\
12 57 20.7 & -17 52 36.4 & 15140 & 33 & Abs & 13.86 & 12\\
\enddata
\tablecomments{The complete version of this table is in the electronic
edition of the Journal.  The printed edition contains only a sample.}
\end{deluxetable}

\end{document}